\documentclass[sigconf]{acmart}

\usepackage{tabularx}
\usepackage{multirow}
\usepackage{caption}
\usepackage{subcaption}

\AtBeginDocument{%
  \providecommand\BibTeX{{%
    \normalfont B\kern-0.5em{\scshape i\kern-0.25em b}\kern-0.8em\TeX}}}

\setcopyright{acmcopyright}

\copyrightyear{2022}
\acmYear{2022}
\setcopyright{acmlicensed}\acmConference[CHI '22]{CHI Conference on Human Factors in Computing Systems}{April 29-May 5, 2022}{New Orleans, LA, USA}
\acmBooktitle{CHI Conference on Human Factors in Computing Systems (CHI '22), April 29-May 5, 2022, New Orleans, LA, USA}
\acmPrice{15.00}
\acmDOI{10.1145/3491102.3501924}
\acmISBN{978-1-4503-9157-3/22/04}

\begin{document}

\title{Remote Co-teaching in Rural Classroom: Current Practices, Impacts, and Challenges}

\author{Siling Guo}
\affiliation{%
  \institution{Lenovo Research}
  \city{Beijing}
  \country{China}
}
\email{lily0516gsl@gmail.com}

\author{Tianchen Sun}
\orcid{0000-0002-8954-958X}
\affiliation{%
  \institution{Lenovo Research}
  \city{Beijing}
  \country{China}}
\email{robertsuntc@gmail.com}

\author{Jiangtao Gong}
\affiliation{%
 \institution{Institute for AI Industry Research, Tsinghua University}
 \institution{Lenovo Research}
 \city{Beijing}
 \country{China}}
 \authornote{Corresponding author}
 \email{gongjiangtao2@gmail.com}
 
\author{Zhicong Lu}
\affiliation{%
  \institution{City University of Hong Kong}
  \city{Hong Kong}
 \country{China}
  }
  \email{zhicong.lu@cityu.edu.hk}

\author{Liuxin Zhang}
\affiliation{%
  \institution{Lenovo Research}
  \city{Beijing}
  \country{China}
}
\email{zhanglx2@lenovo.com}

\author{Qianying Wang}
\affiliation{%
  \institution{Lenovo Research}
  \city{Beijing}
  \country{China}
}
\email{wangqya@lenovo.com}

\renewcommand{\shortauthors}{Guo, et al.}

\begin{abstract}
  The shortage of high-quality teachers is one of the biggest educational problems faced by underdeveloped areas. With the development of information and communication technologies (ICTs), China has begun a remote co-teaching intervention program using ICTs for rural classes, forming a unique ``co-teaching classroom''. We conducted semi-structured interviews with nine remote urban teachers and twelve local rural teachers. We identified the remote co-teaching classes' standard practices and co-teachers' collaborative work process. We also found that remote teachers' high-quality class directly impacted local teachers and students. Furthermore, interestingly, local teachers were also actively involved in making indirect impacts on their students by deeply coordinating with remote teachers and adapting the resources offered by the remote teachers. We conclude by summarizing and discussing the challenges faced by teachers, lessons learned from the current program, and related design implications to achieve a more adaptive and sustainable ICT4D program design.
\end{abstract}



\maketitle

\section{Introduction}
Educational inequality has been a matter of significant concern globally~\cite{wide,burroughs2019teacher}. 
The irrational distribution of teacher quality across advantaged and disadvantaged students has been revealed as a typical example of educational inequality. 
Research has demonstrated that disadvantaged students, indicated by most disadvantage factors (e.g., low socioeconomic status and underrepresented minority), usually have low-quality teachers, measured by nearly every measure of teacher quality (e.g., experience, degree, and even value added) \cite{burroughs2019teacher,schulte2019teacher,pena2016supporting,goldhaber2015uneven,clotfelter2005teaches,lankford2002teacher}.

Gaps in teacher quality between urban and rural areas in China have also attracted much attention
~\cite{xuehui2018teacher}. 
To address the issue of unequal distribution of teacher quality, 
China launched a remote co-teaching program as an information and communication technology used for development/education (ICT4D/E) program in 2013. 
In the program, remote teachers from urban settings typically live stream their classes and rural students watch and interact with the remote teachers via online learning software on the large-size screen in their classrooms. Simultaneously, a local teacher will be present in rural classrooms and assist when needed. 
These remote co-teaching programs are aimed at providing rural students with quality education and enabling rural students to interact with both local teachers and remote high-quality urban teachers \cite{Wang2019,Gu2011}. 
It has been nearly a decade since the first remote co-teaching program was put into practice. To the best of the authors' knowledge, there have not been any empirical studies investigating the current status of the programs in terms of involved teachers' experiences and practices, perceived impacts, and concerns or challenges. 

Monitoring and evaluating ICT4E programs are significant as there can be unexpected problems with the initial design and application of the programs that need to be addressed along the way. For example, for \textit{One Laptop per Child} (OLPC), one of the largest ICT4E interventions, some fieldwork evaluation revealed that the actual practice in Paraguay was usually not program-centric or education-oriented as the program designers hoped but instead leisure-focused \cite{ames2016learning}. 
Therefore, one important take-away from previous ICT4E projects is to rigorously evaluate the benefits of the programs (e.g, through pilot projects \cite{asabere2003encouraging}) before massive investment and large-scale implementation \cite{warschauer2010can}.
In addition, past literature has also pointed out some common areas for improvement in Chinese ICT4E projects, including 1) prioritizing hardware and construction over software and application and 2) implementing one-way information flow from urban (designers) to rural areas \cite{schulte2015dis}. Therefore, understanding the current practices, impact, and challenges of the remote co-teaching programs is important for improving current programs and designing future ICT4D/E programs. An evaluation framework this paper utilizes was proposed by \cite{mumtaz2000factors} and the framework contains factors at school-, resource-, and teacher-levels that might influence the effects of ICT4E projects, especially school-based projects.

In this paper, we explored both local and remote teachers' experience with participating in different remote co-teaching programs. In-depth semi-structured interviews were conducted with each teacher participant, and the transcripts were analyzed following the thematic analysis method. We found that remote teachers tended to assume more responsibilities before and during class (e.g., preparing lessons and lecturing) while local teachers engaged more after the co-teaching class. The patterns of practices also varied by different types of programs. In addition to the direct positive impact of remote teachers on local students and teachers, local teachers who took the initiative and adapted the input resources brought the indirect positive impact of the remote co-teaching system. Although we believe that overall the remote co-teaching program is a successful intervention using ICTs to improve rural education in China, there are still areas for improvement. Teachers faced challenges brought by unclear instruction from the programs, lack of resources and incentives, and the challenging nature of remote collaboration. 
Thus, our main contributions are three folds: i) 
identified the current practices of the large ICT4D/E interventions in China---remote co-teaching programs, ii) analyzed both direct impact brought from remote teachers and indirect impact from local teachers of the programs, and iii) revealed challenges of the programs and discussed design implications of a more adaptive and sustainable ICT4D/E program.

\section{Related Work}
Three areas of research were reviewed. Firstly, we introduce past ICT4E/D projects aiming at reducing educational inequality. Secondly, we discuss the similarities and differences between traditional co-teaching and current remote co-teaching. Last, since remote co-teaching is a form of remote collaboration, we examine technology-mediated remote collaboration and its challenges.

\subsection{ICT4D/E Projects to Reduce Educational Inequality}
Educational inequalities and divides exist both across and within countries.
For example, in China, with one of the largest urban and rural gap around the globe \cite{yang2007urban}, cumulative education achievement gaps exist between urban and rural students \cite{zhang2018teachers}. Urban students' scores in primary subjects (e.g., Mathematics and English) are significantly higher than the scores of rural students \cite{zhang2018teachers}. One important aspect of educational inequalities is the uneven distribution of teacher resources---underdeveloped areas are more likely to lack high-quality teachers than privileged areas \cite{burroughs2019teacher,xuehui2018teacher}. Gaps in teacher quality could lead to achievement differences between advantaged and disadvantaged students \cite{burroughs2019teacher,zhang2018teachers}. In China, urban teachers have attained higher levels of education and have higher professional ranks in urban schools \cite{wang2012social}. 

One widely discussed solution to educational inequality is to use ICTs \cite{schulte2019teacher,wagner2018technology}. For example, OLPC and \textit{Hole in the Wall} are two iconic ICT4E projects \cite{wagner2018technology}. The OLPC project intends to "empower the world's poorest children through education" and the program provides easy-to-use laptops with designed content and software \cite{OLPC}. In the \textit{Hole in the Wall} project, staged computer-equipped kiosks allowed children to learn and explore on their own \cite{HIWEL}. These ICT4E projects have received few rigorous evaluations to measure their impacts \cite{wagner2018technology}. 
Moreover, according to the existing evaluation of the OLPC project, the project yielded mixed results: the widely distributed laptops had a positive influence on children's cognitive skills, but did not improve children's learning \cite{cristia2017technology}. Furthermore, without structured guidance, children tended to use the laptops only for leisure activities \cite{ames2016learning}. Therefore, for future ICT4E projects, a careful analysis, planning, and evaluation before large-scale investment is necessary \cite{warschauer2010can}. One way to systematically evaluate program efforts is to use a framework that summarizes the factors contributing to the success or failure of ICT4E programs. One of the most widely-used frameworks that targets school and teacher level factors influencing ICT use was proposed by Mumtaz \cite{mumtaz2000factors}. The framework calls attention to schools' organization of time, supportive network for teachers to learn and reflect, training opportunities for teachers, as well as teachers' and schools' beliefs about changes brought by ICTs. Using this framework \cite{mumtaz2000factors}, the lessons learned from the OLPC project highlighted the importance of incorporating pedagogical elements into the device-based approach \cite{wagner2018technology}, as well as providing adequate facilities and skill training for teachers and school staff.

The remote co-teaching program does not take a device-based approach. Instead, it uses ICTs as communication, collaboration, and online course delivery tools to bring quality education to rural areas. The program design accords with the current educational context of rural China. Most rural schools have been equipped with essential ICTs (e.g., projectors and computers) \cite{xuehui2018teacher}. Therefore, current rural education needs effective ways to use these ICTs to reduce the urban-rural education gap in China.
Given little empirical research on effectiveness of the remote co-teaching program and the importance of monitoring and evaluation, we intended to investigate the impact of the current ICT4E program in China.



\subsection{Traditional Co-teaching and Remote Co-teaching}
The most widely accepted definition of co-teaching is ``two or more professionals delivering substantive instruction to a diverse, or blended, group of students in a single physical space” \cite{cook1995co} (p. 2). The current remote co-teaching model meets the definition. Although the remote teacher is not physically located in the local classroom, the two teachers can still collaboratively deliver the content to the same group of students in the class via online platforms.

Co-teaching has many variations as the actual practices need to be adjusted according to different students', schools' and teacher' characteristics \cite{cook1995co}. Four main application types of co-teaching approach are: 1) supportive teaching---one teacher is in the lead role, and the other provides support, 2) parallel teaching---co-teachers work with different student groups simultaneously in the same class, 3) complementary teaching---one teacher teaches the content while the other helps with access, and 4) team teaching---two teachers equally shared all the responsibilities of teaching \cite{villa2013guide}. Although the types and practices of co-teaching may vary,  co-teachers are encouraged to collaborate in all facets of the educational process to achieve the greatest impact \cite{cook1995co}. Important elements of a successful co-teaching experience include shared goals, positive interdependence, monitoring of the progress, and enough systemic supports \cite{villa2013guide}. These elements can also be helpful when we evaluate the remote co-teaching classes.

Due to the COVID-19 pandemic, recent work has discussed traditional co-teaching models in remote and hybrid school environments \cite{barron2021co,chizhik2020making}. These studies highlighted the role of technology to facilitate in-class small-group discussion and to promote virtual collaboration between co-teachers \cite{barron2021co,chizhik2020making}. Nevertheless, the remote co-teaching program is different from the aforementioned co-teaching in an online environment in various ways. For example, the students, either in local or remote contexts, are grouped together in class instead of joining separately via online software. Rural teachers and students are also present in the same classroom. Hence, offline communication among students and between rural teachers and students is an important factor for the program. In addition, this remote co-teaching is the first time for most co-teachers to collaborate. This situation is distinct from virtual co-teaching during COVID-19 where co-teachers have worked together before and transferred their work together to online collaboration. Therefore, the current study intends to enrich the body of research regarding co-teaching and look at how ICTs might uniquely change the process in the remote co-teaching programs.

\subsection{Remote Collaborative Work}
Despite the teaching skills that co-teachers should acquire, researchers also concluded that collaboration skills were necessary for an effective and successful co-teaching \cite{friend2010co, scruggs2007co}. Such collaboration does not only apply to the classroom setting, but also applies to all contemporary school endeavors \cite{friend2010co}.
Unlike the traditional co-teaching collaboration, remote co-teaching collaboration releases the constraint that co-teachers should be physically co-located to each other. Computer-mediated communication technologies allow collaborators to work together even if they are geographically distributed around the globe \cite{kraut2002understanding,hinds2003out,cramton2001mutual}. With the increasing implementation of remote collaboration in work environments, researchers have studied factors influencing performance and conflicts in remote collaboration. 

Some of the elements required for an effective co-locate collaboration are also mandatory for an effective remote collaboration, namely frequent communication, consensus, and power equality \cite{kraut2002understanding, shah2014collaborative, cramton2001mutual, scruggs2007co}. Noticeably, it is more challenging to control these factors in a remote setting. 
Communication is a tunnel that conveys information and maintain relationships to support collaboration \cite{cramer2015couples, shah2014collaborative, agafonova2018sexism}. However, when collaborators incorporate effective tools for communication, the frequency and efficiency in communication seem more significant for a successful remote collaboration. That is, when people have issues that need immediate actions, co-locate partners would respond promptly, but remote partners would possibly not respond in time \cite{olson2002currently}. In addition, people communicate with each other to build trust, and further construct consensus \cite{hillman2019have, rocco1998trust}. Consensus, also known as the common ground and mutual knowledge, is critical for remote team collaboration \cite{chang2017alpharead, kim2020consensus}. For instance, team members with various backgrounds and lack communication may fail to construct a consensus towards their understandings of the contextual information and the importance of a task. They would experience conflicts on how they worked together to accomplish the tasks \cite{cramton2001mutual}. Compared with co-locate collaboration, people working remotely would become more difficult to construct trust and hence result in the impossibility of reaching consensus \cite{rocco1998trust}. Moreover, co-workers may have conflicts if power is distributed unequally within teams \cite{shah2014collaborative}. People find it difficult to collaborate if their seasoned partners hardly share responsibility for conducting the tasks \cite{kirkpatrick2020coaching}. Consequently, researchers reported that remote collaboration was less productive than co-located collaboration \cite{olson2002currently}, and people encountered more unaddressed conflicts in remote collaborative work \cite{armstrong1995managing, hinds2003out}. 

Extensive research has been conducted to study remote collaboration and communication in the HCI fields. Researchers have investigated the computer-mediated communication tools in teachers' perspectives focusing on the outside-the-classrooms communication between teachers and parents \cite{hillman2019have, gong2021all}. But few have studied teachers' collaboration and communication in the co-teaching setting. In addition to the teachers' collaboration and communication, researchers also examined and discussed remote collaboration and communication in a variety of topics. In general, the factors that were studied were either technology-oriented (i.e., verbal and non-verbal cues \cite{zhou2017goodbye}, choice of communication channels \cite{cramer2015couples}, and novel designs of collaborative tools \cite{kim2020consensus, chang2017alpharead}, etc.), or related to collaborators' trait and characteristics (i.e., trust \cite{rocco1998trust}, leadership \cite{luther2008leadership}, and gender \cite{agafonova2018sexism}, etc.). We note that the existing findings were limited to specific domains and whether these findings could be generalized to the co-teaching environment remains unclear. In this paper, we seek to fill the research gap to discover the challenges that co-teachers experienced in remote collaboration. 

\begin{figure*}[t]
    \centering
    \begin{subfigure}[b]{0.49\textwidth}
        \centering
        \includegraphics[width=\textwidth]{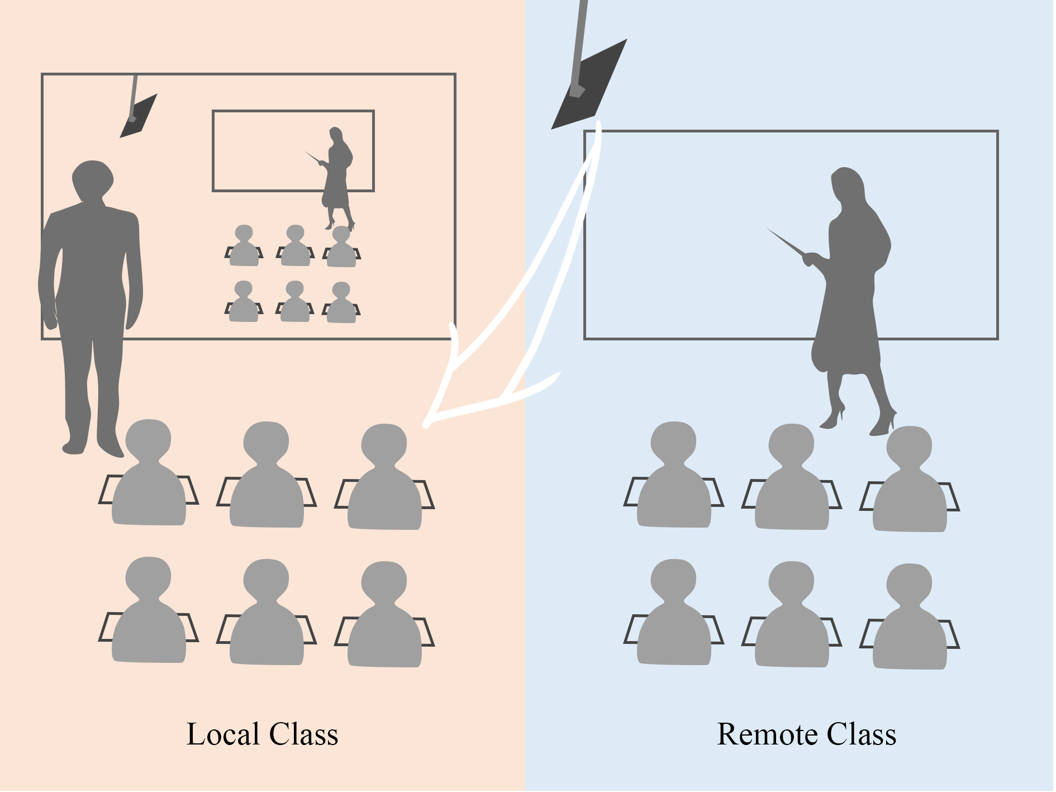}
        \caption{C:C co-teaching program}
        \label{fig:C2C}
    \end{subfigure}
    \hfill
    \begin{subfigure}[b]{0.49\textwidth}
        \centering
        \includegraphics[width=\textwidth]{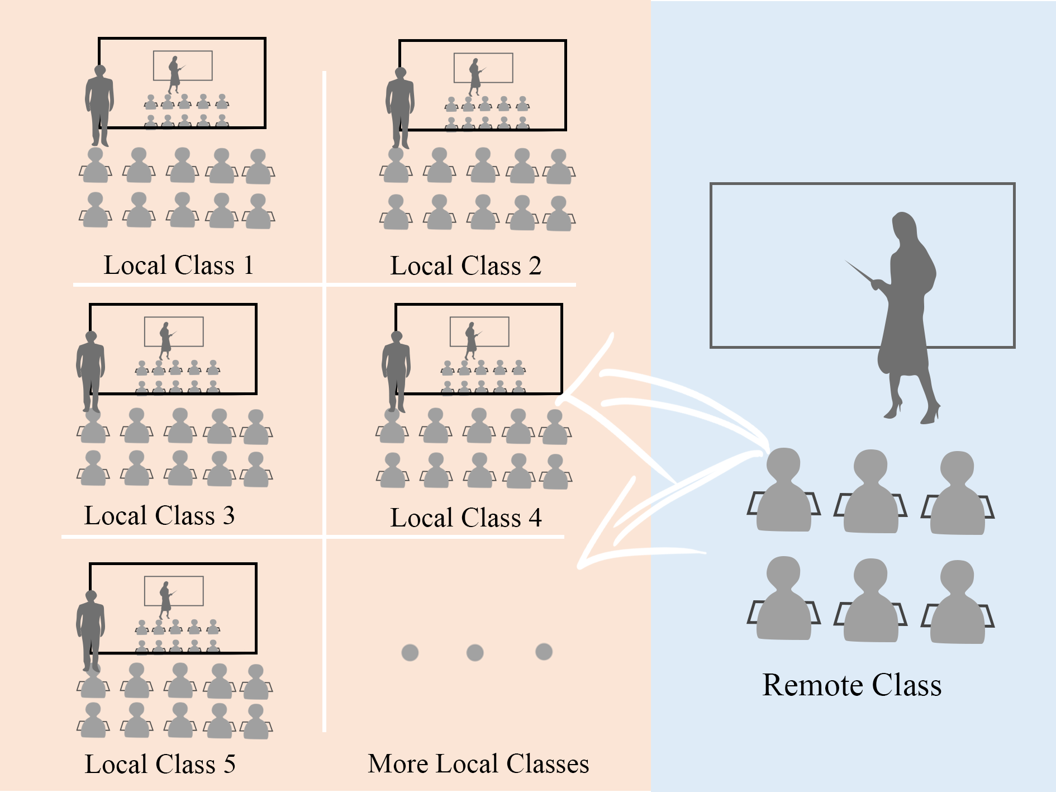}
        \caption{C:MC co-teaching program}
        \label{fig:C2Cs}
    \end{subfigure}
    \begin{subfigure}[b]{0.49\textwidth}
        \centering
        \includegraphics[width=\textwidth]{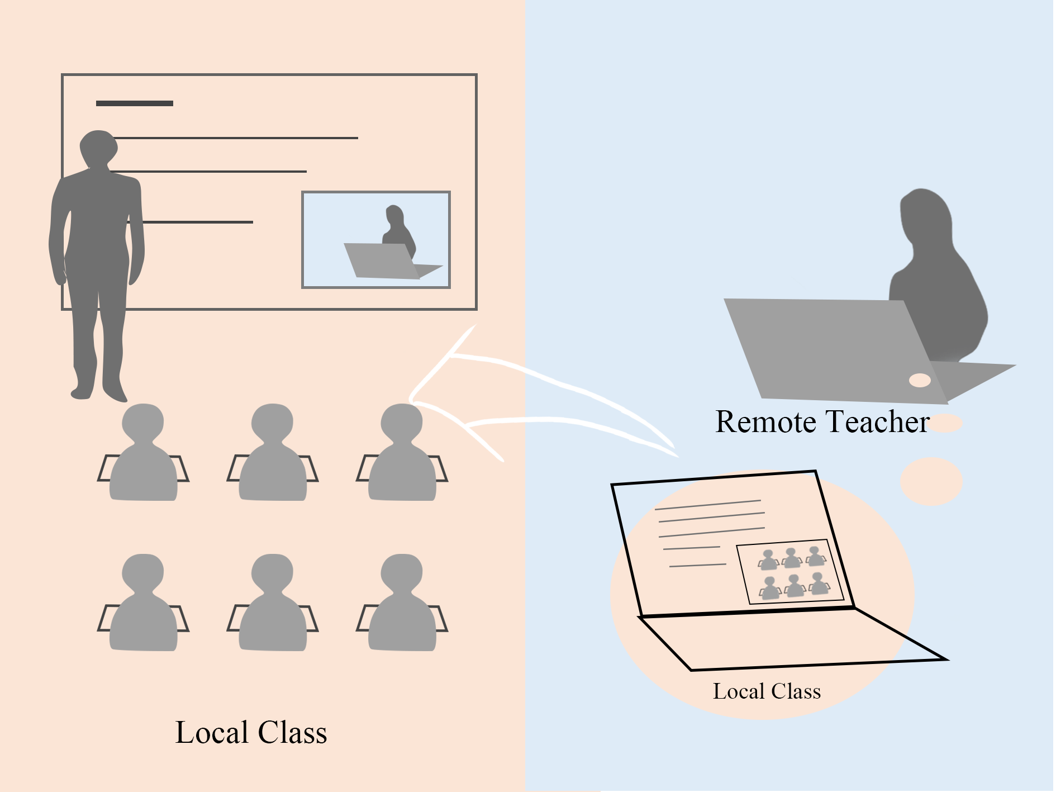}
        \caption{1:C co-teaching program}
        \label{fig:One2C}
    \end{subfigure}
    \hfill
    \begin{subfigure}[b]{0.49\textwidth}
        \centering
        \includegraphics[width=\textwidth]{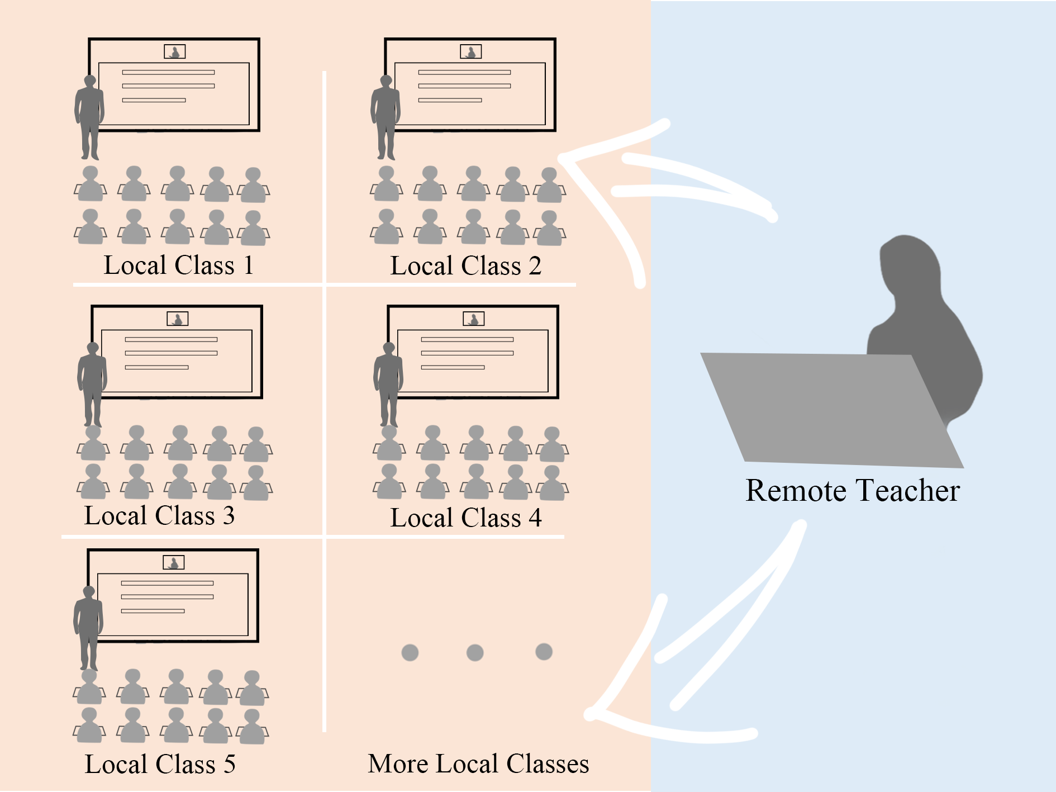}
        \caption{1:MC co-teaching program}
        \label{fig:One2Cs}
    \end{subfigure}
    \caption{Sketches for the four kinds of existing remote co-teaching programs}
    \label{fig:sketches}
\end{figure*}

\section{Method}
\subsection{Research Context}
\subsubsection{Rural-urban Teacher Quality Gap in China}
China has a clear urban-rural divide with a significant income inequality \cite{shi2010re}. In 2020, the ratio of average per capita disposable income in urban to rural China was 2.56 \cite{stats.gov}. \emph{Hukou} system, a household registration system in China, "has created a \emph{de jure} rural-urban divide on top of the \emph{de facto} rural-urban divide \cite{song2019hukou}." The system gives priority to urban households in most life opportunities, such as education, job, housing, and health care \cite{song2019hukou}. 

The resource and opportunity gaps between urban and rural China have led to one-way rural-urban mobility for teachers. At least 7.6\% of teachers transfer out of rural schools each year \cite{xuehui2018teacher}. Such mobility of teachers has created a significantly unequal distribution of high-quality teachers with higher educational levels and senior professional grades across urban and rural China \cite{xuehui2018teacher}. 
In addition to the differences in teachers' degree level, the quality of teacher training programs for Chinese urban and rural teachers also results in the differences in teacher quality \cite{zhang2018teachers}. Urban teachers usually graduate from selective universities, which only admit students who achieve a certain high level on the national college entrance exam \cite{zhang2018teachers}. In contrast, rural teachers do not undergo such a competitive and rigorous selection process and most rural teachers do not have formal teacher training in the subject matter they teach \cite{zhang2018teachers}.
Teacher quality has been supported by past research as an important factor influencing student achievement \cite{chu2015impact,rothstein2010teacher,zhang2018teachers}. Therefore, it is of great significance to address the shortage of qualified teachers in rural schools and gaps between urban and rural teachers in China.

\begin{table*}[t]
\caption{Categories of different remote co-teaching programs}
\label{tab:categories}
\begin{tabularx}{\linewidth}{XXX}
\toprule
Delivery Format/Adaptability Degree
 & Low Adaptability                       & High Adaptability                  \\
 \midrule
Live Streaming      & C:C-live \& 1:MC                     & \textbf{1:C \& C:C}              \\
Pre-recorded        & C:MC-recorded (original)               & \textbf{C:MC-recorded (adapted)}   \\ \hline
Notes              & Unsatisfying results and fewer samples & Promising results and more samples\\ \bottomrule
\label{Categories}
\end{tabularx}
\end{table*}

\subsubsection{Types of Remote Co-teaching Program}
The Remote co-teaching program, which is a relatively new Chinese ICT4E project, utilizes a co-teaching approach in the sense that it involves "bringing" a remote teacher from an urban setting into a rural classroom. While traditional remote learning is one student accessing live streaming or pre-recorded sessions at his or her own place alone, the remote co-teaching projects have extended the audience to one or many classes of students as a group as well as their teachers. To summarize, four kinds of such remote co-teaching programs have existed: 1) class to class (C:C), as shown in \autoref{fig:C2C}, meaning the live streaming sessions or the pre-recorded videos of the urban classroom were directly transferred to the targeted rural classroom; 2) class to multiple classes (C:MC), as shown in \autoref{fig:C2Cs}, referring to those of the urban classroom were transferred to many rural classrooms across the country; 3) one teacher to one classroom (1:C), as shown in \autoref{fig:One2C}, meaning a remote teacher teaches one rural classroom via online class software without remote students on the teacher's end; and 4) one teacher to multiple classes (1:MC), as shown in \autoref{fig:One2Cs}, which is one teacher delivering the class to at least tens of classrooms all over the country. 

\subsubsection{Organizations of Remote Co-teaching Programs}
The four kinds of remote co-teaching programs were typically organized by different sectors: the city's education bureau, top urban public schools, non-government organizations (NGOs), and after-school training institutions, respectively. The C:C programs were usually organized by the local city's bureau of education. The education bureau would partner rural schools with urban schools in the district, and thus, rural schools would receive remote co-teaching classes taught by the teachers from their counterpart support schools. The partnering urban schools usually select teachers with ample experience (e.g.,subject leader teachers) or those with the highest educational backgrounds (e.g., prestigious universities). C:MC programs were typically organized by top urban schools in China. These schools collaborated with several foundations to offer such remote co-teaching programs as experimental ICT4E projects. They would offer either live sessions or pre-recorded videos of those selected teachers' classes to rural schools which joined in the projects. Similar to the selection of the remote teachers in C:C programs, these top schools also selected experienced teachers or teachers with excellent educational backgrounds. Typically, tens or even hundreds of rural schools would have access to these remote classes, either listening to live classes each day together or utilizing the recorded videos covering whole academic years' classes. Some NGOs in China offered the 1:C type of remote co-teaching programs. They would recruit and train college students as remote teachers to deliver online classes to rural students. Although these college students may not have extensive working experience, they all came from prestigious universities and  received at least one semester of training before conducting co-teaching classes. There was usually one liaison in the local school who was responsible for applying to the NGOs for needed courses and arranging pairs of remote volunteer teachers and local assisting teachers. 1:MC programs were generally organized by after-school organizations. The classes were collaborated between one expert teacher teaching in the live streaming studio and many assisting teachers in different local classrooms in different cities.

\subsubsection{Frequencies and Courses of Remote Co-teaching Programs}
In addition to hosting organizations, these remote co-teaching programs were also different in their frequencies and offered courses. The C:MC program was most frequent as it covered mathematics classes for the whole academic year, which took place once each day. The 1:C and 1:MC programs were less frequent, and they would have weekly classes. The classes of the C:C program were the least frequent, with at most one class per month. Moreover, they did not happen on a regular schedule and usually came with short notice (e.g., at most one week in advance) for both urban and rural school teachers. In terms of courses, 1:C and 1:MC programs targeted both exam-orientated subjects (e.g., English and Mathematics) and extracurricular courses (e.g., Arts and Information Security). C:C and C:MC programs, however, only covered exam-orientated courses. 

\subsubsection{Focus of the Current Study}
As shown in \autoref{tab:categories}, these remote co-teaching programs varied in course delivery formats (\textbf{Live Streaming} and \textbf{Pre-recorded}) and the degree of adaptability of the teaching content for local students (\textbf{High Adaptability}: being different for different classrooms and \textbf{Low Adaptability}: being the same across all classrooms). However, according to our interviews, the number of low adaptability programs has decreased or switched to high adaptability ones due to their unsatisfying results. Thus, the focus of the current paper was on the practices, perceived benefits, and challenges that teachers in those high adaptability programs experienced. Therefore, most of interviewees were from the 1:C program, the C:C program, and the C:MC-recorded program.

\subsection{Interview Protocol}
In order to gain a deep understanding of teachers' experience in remote co-teaching programs, 
the semi-structured interviews with further in-depth probing and detailed inquiry~\cite{rubin2011qualitative} were conducted remotely using video or audio calls in June, July, and August 2021.

\subsubsection{Recruitment}
We recruited 21 participants mainly via a snowball sampling method. We connected teachers who we already knew as members of such remote co-teaching programs or organizations via personal and professional networks to invite them to participate in our interviews or get referrals for qualified participants.

Participants' age ranged from 20 to 48 years old. The participants were seven male teachers and 14 female teachers. Their years of teaching experience ranged from 0.5 years to 28 years. Five of them are middle school teachers, and the others are primary school teachers. Their main teaching subjects cover Chinese, Mathematics, English, Arts, and Chemistry. Out of the 21 participants, 12 were local teachers from eight different schools in six different provinces of China. Nine were remote teachers from five different schools in four provinces of China. These participants covered both sides of teachers from all the four main types of co-teaching programs mentioned in Section 3.1. \autoref{demographics} presents more detailed demographic information for each interviewee.

\begin{table*}[ht]
\caption{Demographic information for interviewees}
\begin{tabular}{llllllll}
\toprule
ID  & Gender & Age & Years of Teaching & Grade(s) & Subject(s)             & Type & Program(s)        \\ \midrule
T1  & Male   & 48  & 28                  & 8        & Mathematics         & Local            & C:MC-Recorded   \\
T2  & Female & 35  & 9                   & 8        & Mathematics         & Local            & C:MC-Recorded   \\
T3  & Male   & 43  & 20                  & 3-6      & English             & Local            & 1:C \& C:C    \\
T4  & Female & 36  & 9                   & 3/5/6    & Art                 & Local            & 1:C             \\
T5  & Female & 46  & 22                  & 3        & Chinese             & Local            & C:C             \\
T6  & Female & 37  & 14                  & 5        & Mathematics         & Local            & C:C             \\
T7  & Female & 37  & 15                  & 2        & Chinese             & Remote           & C:C             \\
T8  & Female & 41  & 12                  & 4        & Chinese             & Remote           & C:C             \\
T9  & Female & 39  & 19                  & 1        & Chinese             & Remote           & C:C             \\
T10 & Female & 42  & 18                  & 2        & Chinese             & Remote           & C:C             \\
T11 & Female & 37  & 14                  & 9        & Mathematics         & Remote           & C:MC-Live       \\
T12 & Male   & 26  & 4                   & 9-12     & Chemistry           & Remote           & 1:MC          \\
T13 & Female & 32  & 8                   & 8        & Mathematics         & Local            & C:MC            \\
T14 & Female & 20  & 0.5                 & 5        & Reading and Writing & Remote           & 1:C           \\
T15 & Female & 20  & 0.5                 & 3        & Art                 & Remote           & 1:C           \\
T16 & Female & 20  & 0.5                 & 4        & English             & Remote           & 1:C           \\
T17 & Male   & 42  & 20                  & 3        & Mathematics         & Local            & 1:C           \\
T18 & Female & 24  & 3                   & 3        & Chinese and English & Local            & 1:C           \\
T19 & Male   & 37  & 15                  & 3        & Chinese             & Local            & 1:C \& 1:MC \\
T20 & Male   & 38  & 12                  & 3        & Chinese and English & Local            & 1:C           \\
T21 & Male   & 40  & 22                  & 4        & Chinese             & Local            & 1:C        \\ \bottomrule    
\label{demographics}
\end{tabular}
\end{table*}

\subsubsection{Procedure}
The participants were interviewed through video conferencing software. All the interviews were audio-recorded. The average length of the interviews was around one hour. The authors designed a checklist to help cover all issues concerning this current study (e.g., practices, teachers' perceptions, and their concerns) during the semi-structured interview, as suggested by qualitative researchers in \cite{berg2012qualitative,alshenqeeti2014interviewing}. The key questions of the interview included "Can you tell us the typical process of a remote co-teaching class, from things to do before the class begins until the class ends?", "What is the impact you think the remote co-teaching program has brought to your classroom, your students, and yourself?", "Did you have any concerns before the start of the remote co-teaching program? If so, were these concerns addressed during the course of the program?" and "Where do you think the current remote co-teaching program can be improved?"

\subsubsection{Data Analysis}
The authors transcribed the interviews verbatim and reviewed the transcripts. An inductive thematic analysis approach was applied when analyzing the data, mainly following the step-by-step guided by \cite{maguire2017doing}. Firstly, a line-by-line open coding of the 21 transcripts was conducted by three researchers. All the codes were reviewed together and agreements were achieved for each code. Next, as suggested by \cite{maguire2017doing}, the codes were reviewed and divided into themes and sub-themes to answer the research questions. The authors then discussed and refined the definition and relationship of the relevant themes to the research questions while rereading the transcripts to confirm the meaning. 

\subsection{Background Materials}
To  better understand actual remote co-teaching classes and scaffold the interview protocol design, we collected different forms of background material (e.g., videos and photos, site visit observation notes, and instant messaging chat history) before and during the interview process. The primary source of information was the classroom recordings of the remote co-teaching classes. Some were from open-source videos on the internet and others were provided by the interviewees or organizers of remote co-teaching programs. Furthermore, one author also conducted a site visit and observed a C:C math class and a 1:C art class in a rural school. The site visit occurred before the interview process and served the purpose of the author gaining familiarity with the class modes of remote co-teaching and collecting background information. In addition, some interviewees also provided their chat history with the collaborating teacher to help authors know more about their remote communication. These materials serve as a complement to the interviews. 

\begin{figure*}[t]
  \centering
  \includegraphics[width=0.8 \linewidth]{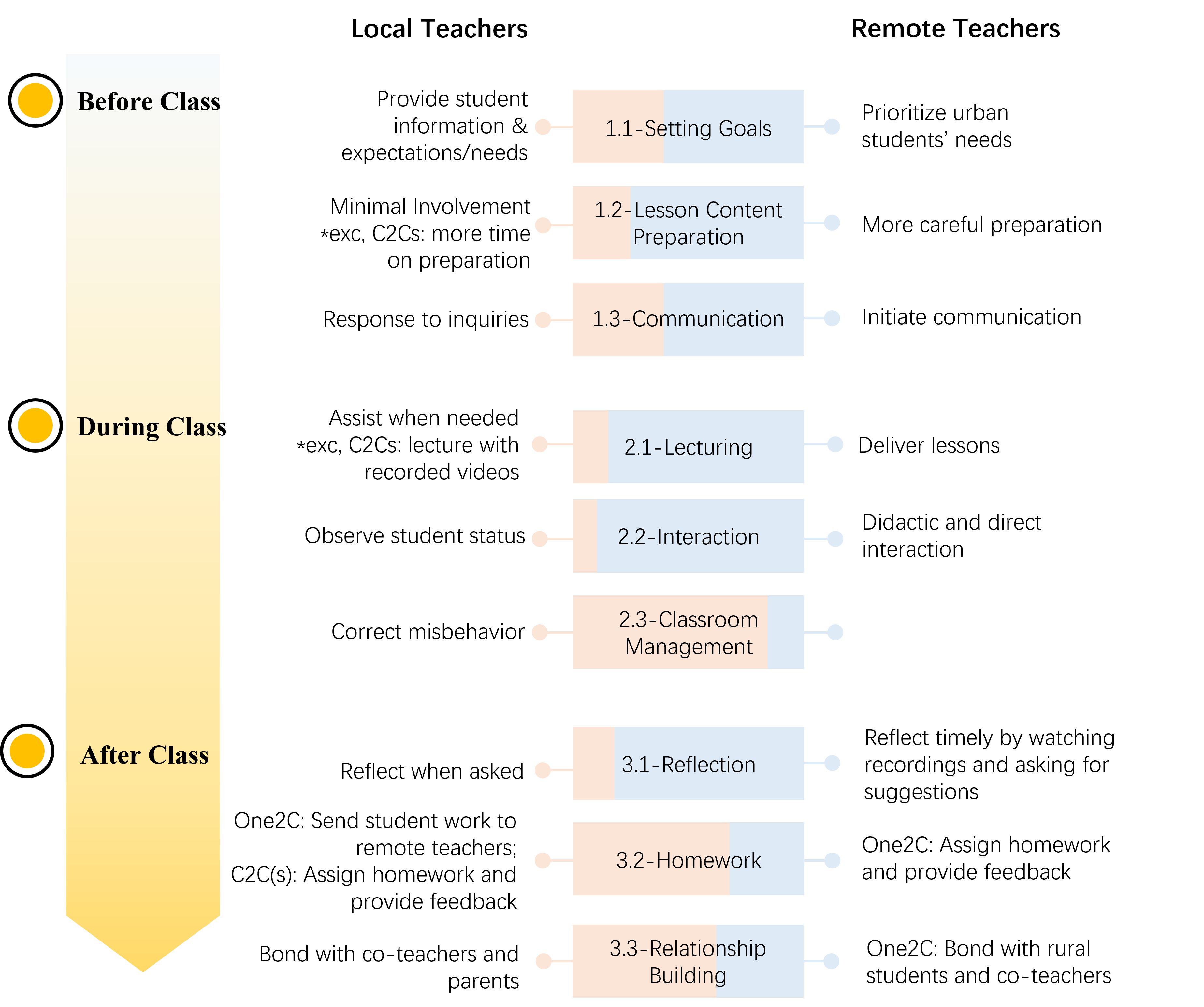}
  \caption[Local and remote teachers' responsibilities]%
  {Local and remote teachers' responsibilities --- \small The color filled in the bars represents the proportion of work done by local and remote teachers.}
  \label{fig:practices}
\end{figure*}

\section{Results}
In this section, we present our results in three parts: 1) current practices of remote co-teaching classes, 2) the perceived positive impact of remote co-teaching programs, and 3) challenges encountered by teachers in the programs.

\subsection{Current Practices}
Different remote co-teaching programs have different focuses and regulations. Therefore, both remote and local teachers tended to assume various responsibilities before, during, and after classes (see the summary of these responsibilities in \autoref{fig:practices}. 
Moreover, the two parties also communicated differently in different programs. Summarizing and comparing the practices of the involved parties in the remote co-teaching programs is of importance as it not only reveals how the programs work, but also points to the degree of cooperation in this unique remote collaborative work.

\subsubsection{Before Class}
Typically, teachers' work before class is to set appropriate teaching goals and to prepare suitable content. A new preparation necessary with remote co-teaching is additional communication between co-teachers. Although the detailed practices varied in different programs, remote teachers took responsibility for setting objectives and preparing lessons before co-teaching. Remote teachers also initiated communication with local teachers to ask for relevant materials (e.g., student information) they needed. Local teachers would follow the plan made by remote teachers and provide the needed information because remote urban teachers were usually considered to be the expert teachers.

In terms of \textbf{setting teaching objectives}, the majority of local teachers helped remote teachers set goals by sharing student information. Generally, local teachers would share their student names and the seating plan through the instant messaging software upon the remote teachers' request. For remote teachers, if they had students on their end, they would set the goals primarily based on their own students (remote students). For example, T11, a remote school teacher, remarked ``I could not prepare the class only for rural students and ignore my own students. Therefore, my lesson plan was still centered around my students." 

While this describes the majority, some local teachers took the initiative to provide more detailed student information and set clear expectations of the course (T18-T21). For example, T19 would mark shy students as well as advanced and underachieved students in the seating plan he sent to the remote teacher. ``I circled out these students, so the remote teacher would know who needs more attention." In addition, four teachers (T18-T21) would proactively communicate their goals of the course with their collaborating remote teachers. T20 said, ``Students have English tests so the textbook has to be covered. Thus, I told him [the remote teacher] to teach the vocabulary and I could teach the readings in the textbooks."

\begin{figure*}[t]
  \centering
  \includegraphics[width=\linewidth]{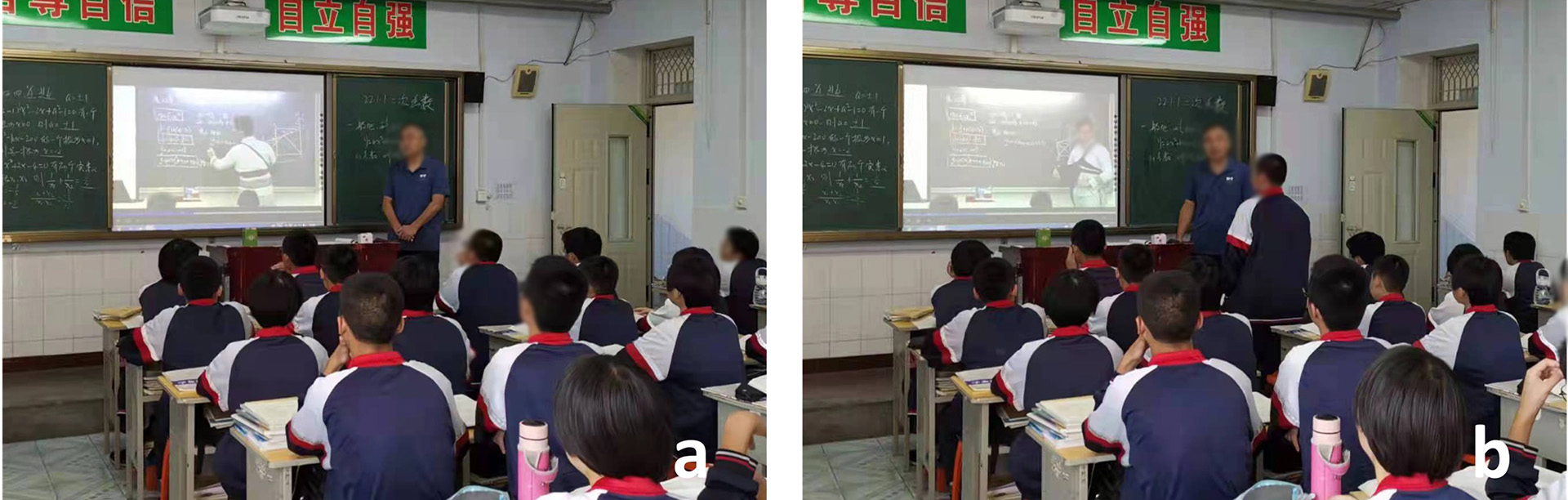}
  \caption{Photos of a C:MC class taken in the local classroom ---\small (a) The local teacher plays the pre-recorded video of the remote teacher's class and he is watching it with the students. (b) The local teacher pauses the video and is asking a student about some questions.}
  \label{fig:c2csrealclass}
\end{figure*}

Most \textbf{lesson content preparation} was accomplished by remote teachers. All remote teachers said that although the routine for preparation of co-teaching classes was similar to that of traditional classes, they prepared co-teaching classes more carefully. T8 said, ``I would write down scripts of the class and then revise them over and over again." T11 would have her mentor review her class slides before each class. As stated by T1, a local teacher, and T9, a remote teacher, some remote schools would even organize a group of teachers to help remote teachers prepare for classes. 

For local teachers, only those in the C:MC-recorded program were responsible for lesson preparation. Moreover, according to T2, the amount of time spent on preparing a remote co-teaching class was two times longer than that of a regular class. Both T2 and T13 suggested that they would watch the pre-recorded videos to mark the time points that they feel need to be paused to provide more explanations for their students. When they became familiar with the videos and picked up the remote teachers' ways of explaining key concepts, they would only watch the video once and then revise the provided class slides during preparation.

\textbf{Communication} between remote and local teachers showed the pattern that the remote teachers initiated and local teachers responded. For pairs of teachers who communicated with each other (mostly in the 1:C program), they used instant messaging and added each other to as a contact (T3-T4, T14-T21). As mentioned earlier, remote teachers would ask their collaborating local teachers for student information. Occasionally, remote teachers would send their lesson plans and class slides to the local teachers for suggestions (T16 \& T18). Some remote teachers would confirm with the local teachers about the availability of resources. Local teachers would usually answer remote teachers' inquiries and respond to their needs (e.g., by printing materials; T14, T16, T17-T21).

\subsubsection{During Class}

The main tasks during class involved lecturing, interacting with students, and classroom management. In remote co-teaching classes, remote teachers were responsible for teaching content via online software and interacting with students, while the local teachers primarily helped manage local classrooms.

\textbf{Lecturing} in a remote co-teaching class is approximately the same as in normal classes for remote teachers. They would introduce topics, explain important concepts, and conduct different activities. The remote teacher's class was live-streamed for rural students. 
Due to the live streaming, T8 and T11 both reported certain differences in lecturing. ``I would be more clear about my instruction so that for rural students, it would still be like listening to a `real' teacher, although I'm only `inside the screen,'" said T8. Similarly, T11 also mentioned that she would choose more demonstrative activities so that it is clear about what is happening both on her end and to the rural students on the other end. 

Most local teachers assisted certain classroom activities. For example, T17 and T21 would help students with in-class exercises and divide students into small teams for group activities. Most local teachers said that they would never intervene during class because they believe it is disrespectful to remote teachers. One exception is that in T19's school, local teachers were asked to intervene whenever they noticed students did not grasp the content. ``If our teachers found many students not understanding a point introduced by the remote teacher, they [local teachers] would ask the remote teacher to repeat," said T19. Moreover, local teachers in T19's school would also write down important points in the class, and use the notes to ask warm-up questions before the next class.

While most local teachers were not involved in lecturing, it was not the case for local teachers in the C:MC-recorded program. These teachers would only use around 10 minute video to introduce important concepts, during which they would pause to provide further explanation or guidance (see class pictures in \autoref{fig:c2csrealclass}). For the rest of time, they would utilize the revised class slides, which were originally provided by the urban school, to deliver lessons.

\begin{figure}[t]
  \centering
  \includegraphics[width=\linewidth]{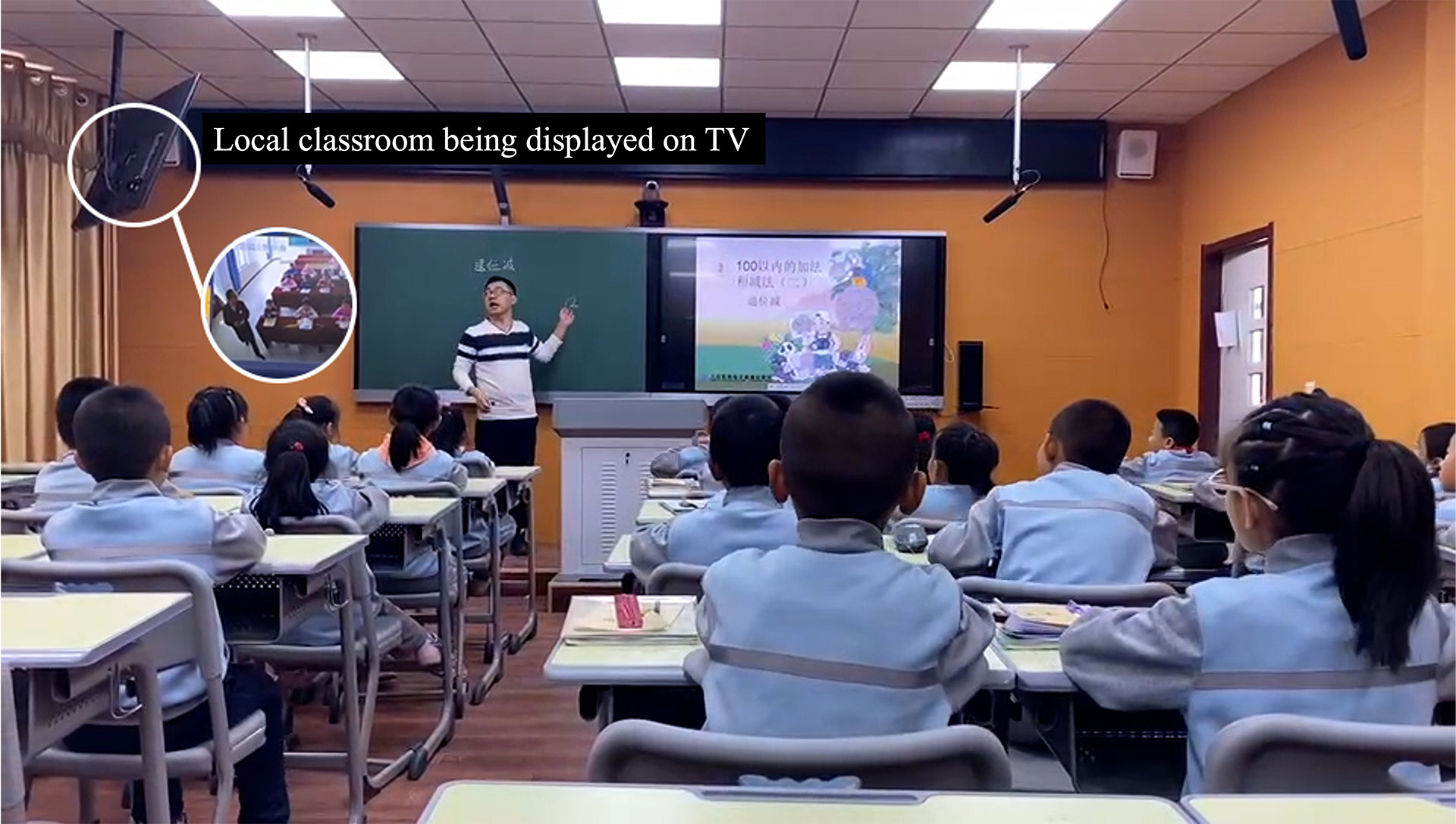}
  \caption{A photo of a C:C class taken in the remote classroom --- \small This is a C:C remote co-teaching class of elementary mathematics. The remote teacher is checking with the local students (whose videos are displayed on the TV) whether they can see the number on the blackboard.}
  \label{fig:c2crealclass}
\end{figure}

\textbf{Interaction with students} showed different patterns between remote and local teachers. Generally, remote teachers directly interacted with students. In 1:C programs, remote teachers would interact with students using the online platform. They asked rural students questions, gave them feedback, and arranged different activities. 
These remote teachers would call students by their names using the seating plan sent by the local teachers. By contrast, remote teachers in the C:C program called students according to where they sat and tended to only ask students who had their hands raised. These teachers interacted with students via a hanging TV that displayed rural students as shown in \autoref{fig:c2crealclass}. Moreover, as these teachers had students physically in the same classroom, they primarily focused on their own students but would try to "switch to the rural school's scenario from time to time," said T7. 

\begin{figure*}[t]
  \centering
  \includegraphics[width=\linewidth]{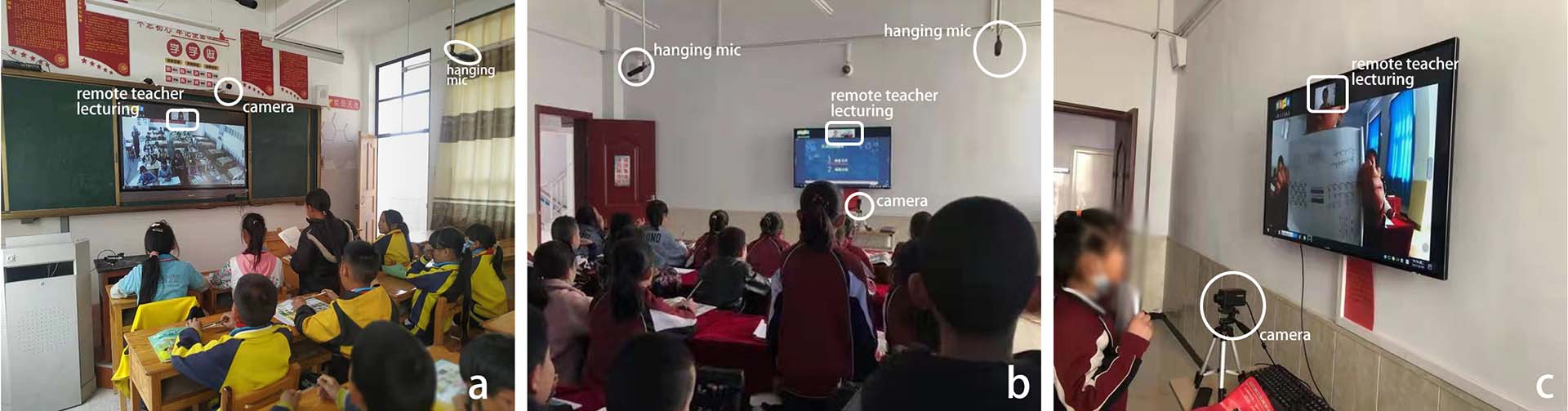}  
  \caption{Photos of 1:C classes (taken by local teachers) --- \small (a) The remote teacher is lecturing and the local students can see the teacher as well as themselves on the screen. (b) The local students can see the class slides and the remote teacher on the screen. One student is answering questions. (c) One student is standing near the camera to show her work to the remote teacher.}
  \label{fig:one2c}
\end{figure*}

Local teachers would observe and monitor their students when local teachers lectured. T1, a local teacher in the C:MC-recorded program, mentioned that when playing the videos of remote teachers' lectures, local teachers could observe students and thus know better about whether students understand the content. Moreover, asked by the organizations, local teachers in 1:MC programs took pictures of class activities and student work as documentation for the class. \autoref{fig:one2c} are pictures of 1:C classes taken by a local teacher.

Most responsibility of \textbf{classroom management} fell on local teachers during remote co-teaching classes. Local teachers would remind students to mind their behavior before class and correct misbehavior during class. As mentioned before, local teachers did not want to disturb the class. Therefore, even when correcting students' behavior, local teachers would avoid using their words but instead remind students with eye contact (T4, T17, T18).  

\subsubsection{After Class}
 For teachers' responsibilities after class, we identified three major categories: lesson reflection, designing homework, and relationship building.

\textbf{Lesson reflection} was usually performed by the lecturers---remote teachers. For remote teachers in the C:MC program, their schools would arrange a team of teachers to observe and reflect the class together. For example, T9 mentioned that other teachers would help point out areas for improvement, such as paying attention to rural students' reactions. Some remote teachers in the 1:C program would review recordings of their classes. "I would focus on whether I asked questions clearly and how students reacted," said T14. Remote teachers in the 1:C program would ask for the local teachers' feedback. When requested by the remote teachers, local teachers would provide their observations and comments.

\textbf{Homework} was designed differently in different programs. In the 1:C program, remote teachers usually assigned homework. Local teachers would help take pictures of finished work and send to the remote teachers. Then, the remote teachers would correct student work and give feedback during the next class. In the C:C and C:MC programs, in most cases, the local teachers were responsible for assigning and correcting assignments. T2, a local teacher in the C:MC-recorded program, said she sometimes assigned the unfinished parts of the remote teachers' lecture recordings as homework. Only one teacher in the C:C program mentioned that she assigned homework to both groups of students, but the homework would be "open-ended.". "After normal classes, I would assign exercises in the exercise book. But in remote co-teaching classes, I would assign some open-ended work that students would be more willing to finish even if they were unsupervised by me," said T7.

\textbf{Relationship building} included strengthening student-teacher relationships, relationships between local and remote teachers, and home-school relationships. For local teachers, they would try to know their collaborating teachers and share remote co-teaching classes with parents. Local teachers became familiar with remote teachers through participating in the same offline activities or following their social media. T1 and T2 would join in professional development programs organized by the urban schools and thus met with the remote teachers. Both T1 and T19 mentioned they invited the remote teachers to the local schools to deliver demonstrative lessons and participate in school activities. Moreover, T18 would look at the remote teacher's posts on social media and congratulate her on her personal achievements. For enhancing relationships with parents, T20 mentioned that teachers in his school shared pictures of remote co-teaching classes with parents. 

From the remote teacher side, they built relationships with students and collaborating teachers. Some remote teachers would purchase prizes for students and deliver them to local teachers (T17, T18, T20). Moreover, some remote teachers would also read their collaborating teachers' posts in social media and form an impression of the local teachers. As mentioned by T16, ``The teacher's posts were mostly about students. I thought this teacher was wonderful, and we did go along well."

\subsection{Positive Impacts}
The positive impacts brought by the remote co-teaching programs are two-fold. First, remote teachers directly impacted local students and local teachers by providing high-quality class and teacher training. Second, some local teachers indirectly impacted the influences of remote teachers on local students through the collaboration with remote teachers.

\subsubsection{Remote Teachers' Direct Impacts}
\begin{figure}[b]
  \centering
  \includegraphics[width=\linewidth]{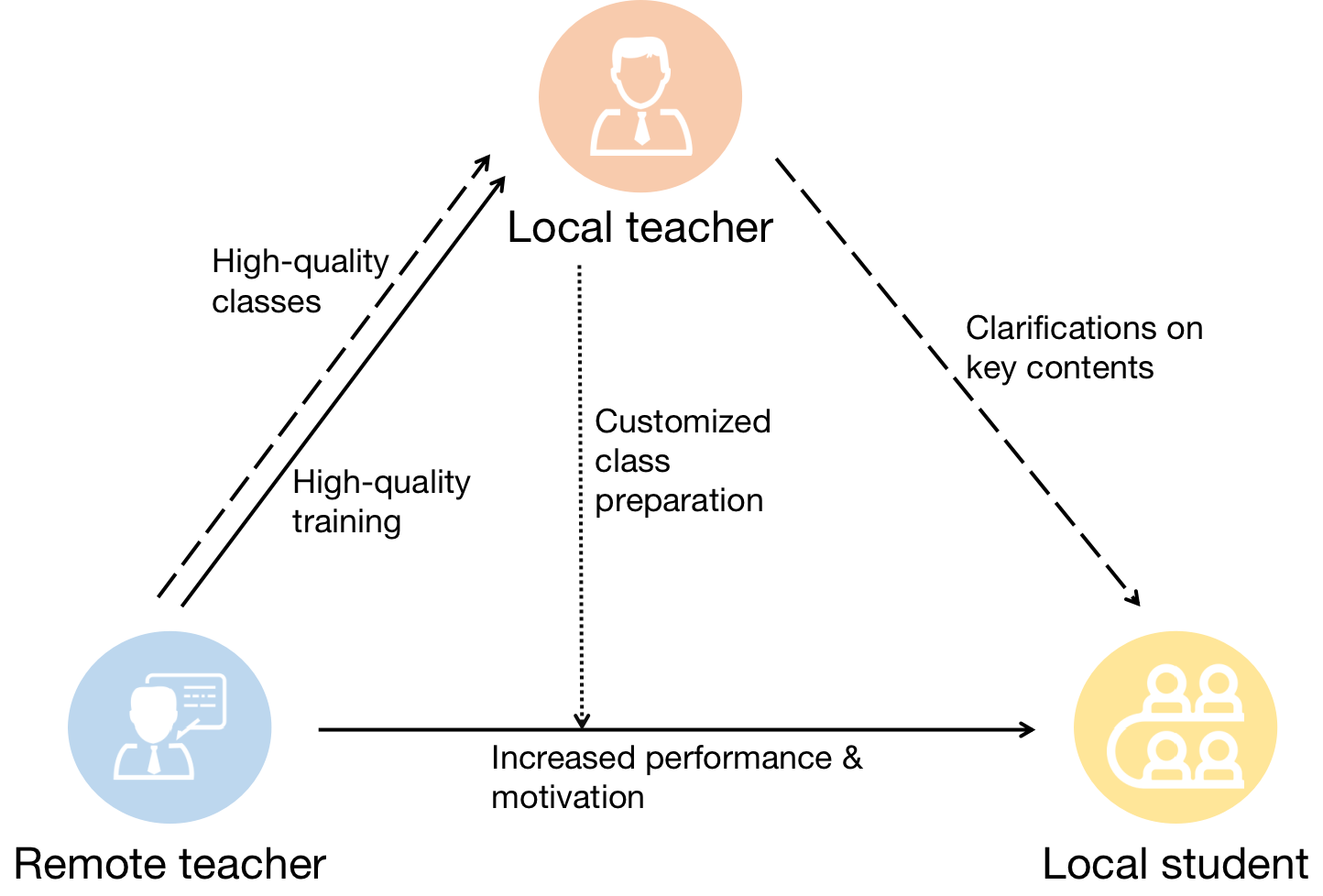} 
  \caption[Positive impacts from teachers]%
  {Positive impacts from teachers---\small Solid lines represent the direct relationships; the dotted line represents the indirect relationship from local teachers as a moderator; and dashed lines represent the indirect relationships from local teachers as a mediator.}
  \label{fig:impact}
\end{figure}

The remote co-teaching classes positively impacted students' learning performances and teachers' teaching skills. As the co-teaching programs designed initially, the most direct impacts on local teachers and local students came from the remote teachers, as shown in \autoref{fig:impact}.

The direct influence on local students included \textbf{improvements in academic performance} and \textbf{enhanced motivation in class}. Multiple local teachers (T1, T2, T18, T20) agreed that local students participating in the programs had broadened their horizons and achieved higher grades. For example, T20 said, 
\begin{quote}
    ``After the semester taught by the remote teacher, this class’ [English] final grades ranked the third in our town. It was a great achievement! However, this semester we did not get English remote teachers. Students’ grades were not as good as the last semester.”
\end{quote}
We believed remote teachers' higher-quality lessons contributed to students' improvements. The quality of classes that local teachers could deliver was limited. Both T18 and T20, who taught in underdeveloped areas, admitted that they were responsible for multiple courses, including which they were not professional at. 

In addition, local students were reported to be more active and motivated because of co-teaching classes. When local teachers turned on the co-teaching software prior to the classes during the break, the students would excitedly greet the remote teachers. During classes, some local students would consider in-class activities as competitions with the remote students, and would actively raise hands to answer questions. Nevertheless, some local teachers (T2, T5, T16) reported that students' high level engagement inspired by remote teachers would only temporarily influence normal classes taught by local teachers. T5 described, ``In co-teaching classes, students were more motivated to answer questions. Students would still be that active for maybe two classes after the co-teaching class. Their level of engagement would gradually decrease.”

The direct influence on local teachers was mainly brought by the \textbf{accesses to high-quality training}. The local teachers used the remote teachers’ instruction as training materials for improving their own teaching skills. As the local teachers continuously participated in the co-teaching programs, they would intentionally or unintentionally mimic the remote teachers’ teaching strategies (e.g., use of animations and attractive contents to raise students’ interests) and teaching styles (e.g., use of languages). The outcomes of the local teachers’ learning from the remote teachers were outstanding (T1, T2, T4, T13, T18, T19). For example, T1 remarked that the gap between grades of the two classes with and without remote teachers became smaller as she began to adopt the remote teachers’ teaching styles. Moreover, T4 indicated that she learned poster painting from the remote teachers' classes. She later applied what she learned in her classes. Local teachers preferred the co-teaching programs to the traditional training sessions because co-teaching resources were long-term while the traditional training sessions were short-term (T1, T2, T13). 

\subsubsection{Local Teachers' Indirect Impacts}
When the co-teaching programs just started, the remote teachers were intended to be the only stakeholder that impacted the classes without additional support from or collaboration with the local teachers. However, after years of practices, the schools discovered that the co-teaching outcomes were not as good as expected if the local teachers remained invisible. The classes were not always successful and the students' grades were not satisfying. Therefore, local teachers began to indirectly impact the co-teaching classes as a moderator and a mediator, as shown in \autoref{fig:impact}.

Local teachers served as a moderator in helping remote teachers \textbf{prepare customized classes for local students} to improve class quality. We noted that the remote teachers sometimes could not prepare classes that were suitable for the local students. For instance, the class contents possibly required prerequisite knowledge that the students did not fully understand, and thus, it would be more difficult for students to catch up in classes. The local teachers would then provide feedback for the remote teachers to help remote teachers know where to improve in the future (T5, T14, T17-T20). Moreover, local teachers would also report the parts in class that students felt bored to the remote teachers. For example, T17 said, ``In my feedback to the remote teachers, I told them about the students’ interests...I also took pictures of the students for the remote teachers, showing them students’ positive reactions to the classes and when they have no interests.”

In addition, local teachers collaborated with remote teachers to designing customized classes. Local teachers were more familiar with the local schools’ resources and their suggestions on the class designs could better allocate their resources, which had also shown great outcomes (T19). For instance, T19 said, 
\begin{quote}
    ``We did not know how to design Painting classes in the past...I told the remote teachers our school had plentiful rock resources so they taught students how to draw on rocks. Now rocks with students’ drawings are everywhere in our campus. Students were very happy to let everyone see and appreciate their work.”
\end{quote}


Local teachers also indirectly impacted the local student-remote teacher relationship as a mediator by \textbf{clarifying key contents in lessons}. Influenced by the classes taught by remote teachers, local teachers could further impact local students and make the co-teaching classes more effective. Unlike 1:C and C:C classes, C:MC classes were still instructed mainly by the local teachers. The local teachers found that students may not fully understand the contents if they played the recorded videos for the entire class because the remote teachers’ pace was too fast. Therefore, some teachers (T2, T13) decided to modify the lecture slides to better adapt to the local students’ backgrounds and use only part of the recorded videos in classes. In addition, the remote teachers' classes sometimes skipped prerequisite knowledge that remote students understood but local students did not acquire. The local teachers (T1, T2, T13) would pause the video and teach the contents by themselves.

Local teacher could monitor and check whether the local students got distracted while remote teachers were instructing. For instance, some local teachers (T2, T6, T13) would focus on assisting the distracted students individually during in-class activities. Another local teacher (T19) would directly interrupt the remote teachers’ class and explain the key points again to the students, as described in section 4.1.2. If the students still had any unresolved questions after class, the local teachers would explain again in their regular classes. T5 mentioned, ``I asked students whether they had any questions after the [co-teaching] classes. I would explain those unclear points again to the students.”

\subsection{Challenges}
Despite the aforementioned benefits, some aspects of the implementations and outcomes of the programs did not meet the designers’ expectations yet. For one, the frequency of remote co-teaching classes was rather low in some programs (T3, T5-T10). For instance, after two years of the C:C program, all the C:C teachers interviewed (T3, T5-T10) claimed that they had experienced fewer than four co-teaching classes. Besides, their whole schools usually scheduled only two to three co-teaching classes per semester for each exam-oriented courses. For another, some remote co-teaching classes' effects on students' performance are mixed. A few local teachers (T5, T13) found local students’ test performances either did not change or even dropped after adopting the C:MC and 1:C programs. 


We believed these unmet goals were due to the following challenges: lack of instruction, training, and rewards for participant teachers from the systems and schools and ineffective remote communication between co-teachers. 

\subsubsection{Lack of Instruction}
Few organizations had regulations for co-teaching classes (T14-T16). Without such instruction, local and remote teachers were often confused about their roles in the co-teaching program. Some teachers (T3, T5-T10) did not even know who the teacher on the other side would be before classes. T4 explained, 
\begin{quote}
    “My school did not tell me what to do. I was only told that there would be a weekly online class. I should collaborate with the [remote] teacher and be in the classroom during class. My school did not ask me to communicate with the [remote] teacher, and there were no requirements on the course contents at all.”
\end{quote}

In addition, many local teachers (T3, T5, T6) did not connect local students with remote teachers well. Local students did not know remote teachers and may get nervous in class. T6 explained, ``Some [local] students from low-income families were self-abased. They rarely raised their hands in class. They became even less active in co-teaching classes because they were not familiar with the [remote] teachers.” Furthermore, remote teachers (T7-T12) did not know backgrounds of local students, which would make it more difficult for the teachers to deliver classes. For instance, T10 commented, ``I did not know how well the [local] students’ studied. I worried that the [local] students would have difficulties understanding the materials if I taught too fast.” 

Besides, we noted that some schools rarely planned the co-teaching classes ahead. Teachers (T3, T5-T10) were usually unaware of the plan until one week before the scheduled class. ``We were only notified about the time [to participate in the co-teaching class]. We just brought students to the classroom then,” T6 said, ``our school might have already contacted the remote school and have scheduled everything but I was only aware of when it would happen.” 

\subsubsection{Lack of Training}
Generally, schools did not provide training to help teachers better deliver a co-teaching class, such as that of interacting with remote students and of using co-teaching software. Therefore, remote teachers (T7-T12) prepared the lessons using the same strategies they used for a regular class and felt that they could not engage local students as effectively. 

Although remote teachers in the I:C program received certain pre-service training, some of their collaborating local teachers (T3, T4, T20, T21) still feel these remote teachers lacked teaching experience. The insufficient teaching experience and the nature of online classes weakened these remote teachers’ control over the class. In addition, the remote teachers were less skilled at preparing the class contents according to students’ backgrounds, interests, and age characteristics. For instance, T4 explained, “when the remote teacher taught students to draw bridges, she talked a lot about the theories, which the third graders were definitely not interested in.” 

In addition, training for technical skills was also necessary because teachers reported they had trouble using video and audio systems in classrooms (T3-T13, T17-T21). However, neither local nor the remote schools equipped teachers with skills to use or trouble-shoot the systems. Almost all the teachers (T3-T13, T17-T21) indicated that they were not familiar with the online platforms and they always required assistance from technicians to turn the software on and trouble-shoot when needed, which led to teachers' lower self-efficacy in delivering co-teaching classes. 

\subsubsection{Lack of Rewards and Stimulation}
As described in section 4.1, some teachers, including remote teachers of the C:C program and local teachers of the C:MC program, had higher workload than usual when participating in the co-teaching programs (T1, T2, T7-T10, T13). Nevertheless,
schools did not provide equivalent rewards or stimulation for teachers’ extra workload, which might be one of the reasons that teachers were not motivated to participate in co-teaching classes. Some teachers (T5, T6, T8, T17, T18) indicated that they received no additional rewards for delivering the co-teaching classes. Others (T7, T9-T11, T13-T16, T19, T20) explained that their participation in the program was related to their annual assessment. T20 explained,
\begin{quote}
    ``There was no way to receive extra points [for assessment], but our points could got deducted. Our principal said if there were no such [co-teaching] programs, we should teach the classes by ourselves. With the programs, we had fewer classes to teach. Why would we get additional points? But if we do not make use of the programs, there would be points deducted.”
\end{quote}

\subsubsection{Unequal Relationships}
In the co-teaching programs, local teachers usually perceived themselves in a lower social and academic position than the remote teachers. Since the co-teachers hardly knew each other before the classes, they would rely on their pre-existing perception of the local/remote teachers and their relationships could become unequal. The 1:C and C:C remote teachers called themselves as the “main instructor,” and they called local teachers “teaching assistants” instead. Multiple teachers (T1, T3, T5, T7, T10-T12) mentioned they believed local teachers were less skilled in teaching than remote teachers. Therefore, remote teachers rarely relied on local teachers’ inputs while preparing and delivering classes. For example, T10 said, ``The local teachers were aging in general and were not familiar with advanced teaching techs. So I took care of all the things, like preparing class contents and slides.”

Such unequal relationships could cause misunderstandings and conflicts. Local teachers (T4, T17, T18) feel uncomfortable to send requests to remote teachers. For instance, T4 mentioned the remote teacher once taught a class similar to a previous normal class but T4 did not tell the remote teacher to change the content. “I did not tell [the remote teacher] because I respected her. She must have put lots of efforts while preparing for the class,” said T4. In addition, T4 always gave the remote teachers full scores in course evaluation system, given that she ``respected" the remote teachers. 
Moreover, T4 repeatedly mentioned because the co-teaching classes took half of her classes but did not cover any textbook-related content, she could not complete her teaching goals. Nevertheless, she apparently never conveyed her concern to the  corresponding remote teacher (T15) because T15 was not aware of this issue. ``The local teacher did not have much requirement for our co-teaching classes. Since their regular classes were generally about the textbook contents, I thought it would be better to teach the students other materials not related to their textbooks,” said T15. 

\subsubsection{High Remote Communication Cost}
We found that co-teachers experienced high communication costs during remote collaboration. Current communication between co-teachers was limited and inefficient. Co-teachers mainly communicated with each other through WeChat, a widely-used instant messaging software in China. It could take teachers a long time to wait for responses, which further negatively impacted co-teachers' collaboration (T4, T15, T17). T17 explained, ``I did not check my phone very often, so I would not see the messages in WeChat. When I saw the messages, it could have already been too late to reply…Maybe because I replied too slow, the remote teacher misunderstood me and we gradually talked less.” 

As a result of the high communication costs, many teachers (T3-T13) were not motivated to contact their collaborating co-teachers out of class. T7 said she was not sure whether the co-teaching class was successful because she did not receive any feedback from the local teachers about the students’ learning outcomes. 
`` 
However, even if she thought feedback was important, T7 was not willing to initiate conversation and ask for feedback from the local teacher. 

\section{Discussion}
\subsection{ICT Programs to Reduce Educational Inequality}
In general, based on participating teachers' experience, the remote co-teaching program in China created a feasible way to use ICTs for educational interventions. We propose two main explanations on why the remote co-teaching program might work and summarize the takeaways for future ICT4E program design.

Firstly, the remote co-teaching program does not rely on the traditional device-based approach but instead allows shifts in pedagogy. The current approach fits with rural Chinese educational contexts as there are already enough educational technology devices in most rural classrooms and what they lack instead is high-quality education resources \cite{xuehui2018teacher}. Using online platforms to deliver high-quality classes, the remote co-teaching classes overcame the major weaknesses of rural education: a shortage of professional teachers and a well-designed curriculum. More specifically, providing high-quality classes in exam-orientated subjects (e.g., Mathematics in C:MC programs) or those that can not be delivered by rural teachers (e.g., Arts and Music in 1:C programs) aligns with the rural schools' needs. According to the factors influencing the effects of ICT4E programs listed in \cite{mumtaz2000factors}, in this case, the rural schools hold the belief that the changes brought by ICTs are necessary, and thus, the schools would actively participate in the programs. 
This approach might shed light on importance of using ICTs to directly intervene school curriculum in underdeveloped areas.

Nevertheless, most of previous ICT4E programs primarily focused on hardware or devices rather than the pedagogical elements \cite{wagner2018technology}. For example, 85\% of funds of a Chinese ICT4E program would go into hardware, while there is no fund for teachers’ training \cite{schulte2015dis}. Moreover, the two iconic ICT4E programs---OLPC and \textit{Hole in the Wall}---also focused on providing facilities. Although this approach could be beneficial by itself as designed by the two programs, incorporating pedagogical elements and targeting schools' and teachers' needs as mentioned in \cite{mumtaz2000factors} might bring more positive results. Therefore, future ICT4E program might consider directly addressing the distortions in curriculum and pedagogy with the help of ICT utilization, implementation, or design.

Secondly, the remote co-teaching program ``spans more than one subsector" \cite{wagner2018technology} (p. 8) and provides support for both rural teachers and students. Past research suggests that when more than one driver out of the three drivers of changes in education quality (resources, incentives, and community participation) were combined in the intervention programs, they are more likely to be effective \cite{masino2016works}. Other exemplary ICT4E projects that targeted two stakeholders and yielded positive results are: 1) the \textit{Bridges to the Future Initiative} in India that intended to promote literacy in both young children and adults \cite{piper2016does}, and 2) the \textit{Kenya Primary Math and Reading} program that provided both student learning materials and teacher training \cite{wagner2010technology}. Consequently, another takeaway for the design of future ICT4E programs is to integrate various drivers that bring change in the quality of education.


\subsection{Towards a More Adaptive and Sustainable ICT4E Program}
If intervention programs are adaptive and sustainable, they can effect changes (e.g., closing educational gaps) even after the programs ended. The current remote co-teaching program has already been viewed by many teachers as continuous and stable resources leading to long-time positive impact. Meanwhile, the program still has potential to be improved towards a more effective intervention.

To achieve a successful ICT4E program, end users' characteristics 
need to be understood adequately and thoroughly \cite{wagner2018technology}. One problem with the C:MC program might be the expert remote teachers' class videos were not tailored to local students. The content might be out of local students' zone of proximal development and thus, students could not learn effectively \cite{vygotsky1978mind}. However, with local teachers' explanation, the content became understandable for local students. The effective adaption performed by the local teachers highlights the importance of understanding local students' needs and the positive influence that local teachers can make in interventions.

In addition, the low frequency of the C:C program might result from the missing shared vision and insufficient incentives to engage in the program. According to the Managing Complex Change Model, \textit{vision} and \text{incentives} are elements required for effective change \cite{villa2000restructuring}. According to the model, when the vision is missing, people will be confused; moreover, when incentives are not enough, people tend to resist the change \cite{villa2000restructuring}. Indeed, most remote teachers in the C:C program considered co-teaching classes as demonstrative classes, and they did not receive any incentives for the long preparation time. As a result, they tended to feel confused about why they needed to collaborate with local teachers and resist a more frequent and regular-basis co-teaching class schedule. Therefore, one way to improve the program is to share the vision with all the stakeholders and provide rational incentives for teachers.

One promising impact of the program we noticed that might lead to sustainable influence is the change of teachers' educational beliefs. Teachers' belief about how the class should be taught is also an important teacher-level factor determining the success of ICT4E program identified in \cite{mumtaz2000factors}. For instance, some local teachers remarked that they ``adopted a whole-child perspective instead of only focusing on student academic performance" (T18 \& T19) and ``became more willing to encourage students and prioritize their mental well-being" (T20 \& T21). For remote teachers, T10 and T16 mentioned that they became to care more about rural education. Past research supports such changes in beliefs and attitudes among teachers are associated with higher possibilities of uptaking changes in practices \cite{guskey1986staff,cook2015supportive}. We believe that the changes in teachers' beliefs brought by the remote co-teaching program have the potential to make the program more adaptive and sustainable.


\subsection{Efficient Remote Collaborative Work}
Consistent with findings of previous remote collaboration studies, the co-teachers experienced high remote communication costs during collaborations, resulting in decreased communication motivation and ineffective collaboration. Studies indicated that collaboration was a time-consuming process, and the collaboration could be unproductive if the task required quick responses and actions \cite{shah2014collaborative}. Such communication challenge was much severe in a remote collaboration setting given that parties working remotely usually lacked an interactive and continuous communication \cite{hinds2002distributed}. 

It is interesting to note, some teachers reported to have successful collaborations with their co-teachers. We found these effective collaborations benefited from frequent and effective communication between co-teachers. 
Some teachers developed their own strategies to build relationships with their co-teachers. For instance, some teachers (T1, T15) used emojis instead of plain texts when communicating, which was in line with what was suggested by \cite{zhou2017goodbye}. 
We suggest future remote co-workers could learn from these successful collaborations to overcome the challenges in communication, for instance, by constructing detailed plans for communication schedules. 


Besides, using the framework by \cite{mumtaz2000factors}, we found there is a need for providing a supportive network that offers technical skill training, reflection and sharing sessions, assistance in remote collaboration for teachers. Our findings demonstrated that insufficient support from the organizations and schools to the teachers, including the instructions and training, rewards and stimulation, and equal relationship building, were significant challenges that prevented the remote co-teaching programs from being more successful. 
We believe that as the schools and related organizations pay more attention to the co-teaching programs and programs become more mature, these challenges could be resolved.

\subsection{Design Implications}
We found the remote co-teaching program, especially the pre-recorded videos used in the C:MC program, could be more adapted into local contexts. Therefore, we suggest future technologies supporting remote co-teaching could utilize intelligent tutoring systems (ITSs) to promote adaptive instruction and learning, such as \cite{rosen2018effects,sottilare2017enhancing}. The recorded videos of remote teachers' classes can be integrated as modular video clips so that local teachers can select to play any video clips based on their needs. The advanced ITSs can also allow local teachers to inform the remote teachers about the local students’ needs. The usability of the ITSs should also be further improved for reduced learning costs. Besides, with fast development of artificial intelligence technology, we can imagine an AI teacher that could integrate teaching styles, strategies, course contents, and communication styles of human teachers in the future. 

Our results also indicated that participating teachers suffered from high communication costs to maintain relationships with co-teachers. Moreover, teachers mostly used instant messaging software to communicate outside classrooms and the video conferencing groupware inside classrooms. However, such communication was mostly work-related and remote teachers could hardly construct emotional bonds with the local students and teachers. Asynchronous communication tools, such as Discord and Slack, could be adopted to enable more flexibility in teachers' communication. Local teachers could share news, grades, fun activities of students, and their own thoughts with the remote teachers in a private session, where instant responses are not expected. Remote teachers could know more about the local students and would develop common ground with the local teachers that would further enhance communication and collaboration qualities. 

In addition, augmented and alternative communication (AAC) technologies could be investigated and adopted in the co-teaching programs to enhance communication and induce participation in remote collaborations \cite{shin2020talkingboogie, mirenda2003toward, hunt2002collaborative}. Replacing complicated sentences with symbols and non-verbal cues allows teachers to quickly express their ideas and dramatically reduces their communication costs \cite{zhou2017goodbye}. These tools can mediate the geographical distance between co-teachers and share each other’s awareness on the collaborative work. In addition, since co-teachers were sometimes not aware of what tasks they should collaborate on, future co-teaching tools can specify a list of tasks required for co-teachers.


\subsection{Limitation and Future Work}
 Firstly, we acknowledge this study is an initial investigation of the remote co-teaching programs with a small sample size recruited from snowball sampling. More generalizable evidence need to be collected.
 Secondly, we note that the indirect impacts from local teachers can be categorized as a moderating effect (the dotted line in \autoref{fig:impact}) that enhanced the positive influences of remote teachers on the local students, and a mediating effect (the dashed line in \autoref{fig:impact}) that mediated the high-quality classes from remote teachers and delivered to the local students with clearer explanations. 
 However, quantitative data from follow-up surveys is necessary to further confirm the robustness of the proposed moderating and mediating effects. 

For future studies, we suggest researchers to more focus on one of the co-teaching programs and examine the strengths, drawbacks, and collaboration of the co-teaching programs more specifically. In addition, researchers can conduct controlled experiments to draw a more rigorous conclusion about the impacts of the co-teaching programs on local students' learning and development.

\section{Conclusion}
In this paper, we presented a qualitative study examining an ICT4E program---a remote co-teaching program in China. We interviewed 21 teachers who had experience in one of the four existing co-teaching programs, namely C:C, C:MC, 1:C, and 1:MC. To the best of the authors' knowledge, this study is the first investigation of the remote co-teaching programs in China. We found that both local and remote teachers involved in these programs were responsible for some tasks during the whole educational process (e.g., before, during, and after classes). In addition, we noted that remote teachers had direct positive impacts on both the local students and local teachers, while local teachers indirectly influenced relationships between local students and remote teachers. Furthermore, though the co-teaching programs have been implemented in some rural and urban areas in China for years, challenges that preclude in-depth collaboration between co-teachers still persist. The challenges included lack of instructions, training, and rewards from the system side as well as unequal relationships and high-quality community cost from the remote collaboration side. The results of the study shed new light on guidelines of designing effective ICT4D/E projects. The finding of this research also provide insights of designing technologies that better support adaption and remote collaboration. We expect this work to contribute to the ICT4D/E research community by providing lessons and inspirations from a relatively long-lasting ICT4E program in China. Our study also contributes to the wider HCI community that the factors identified as enhancing or hindering an effective remote collaboration in the remote co-teaching field could also be applied to more generalized HCI fields. 

\begin{acks}
This work was supported by the China National Key Research and Development Plan under Grant No. 2020YFF0305405. 
We thank the non-governmental organization \textit{Tong Nian Yi Ke} \footnote{\url{https://www.tn1k.cn}} for generously providing the resources of their 1:C co-teaching program. We sincerely appreciate all the participant teachers' kind sharing of their experience, thoughts, and class pictures.
\end{acks}

\bibliographystyle{ACM-Reference-Format}
\bibliography{sample-base}

\end{document}